\documentclass[prd,twocolumn,letterpaper,superscriptaddress,reprint]{revtex4}

\usepackage{graphicx}
\usepackage{pslatex}
\usepackage{amsmath}
\usepackage{amsthm}
\usepackage{amsbsy}

\newcommand{\beq}{\begin{equation}}
\newcommand{\eeq}{\end{equation}}

\newcommand{\atUH}{Dept. of Physics \& Astronomy, Univ. of Hawaii, Manoa, HI 96822.}

\begin{document}

\title{Antiquark nuggets as dark matter: New constraints and detection prospects}

\author{Peter~W.~Gorham}
\affiliation{\atUH}

\begin{abstract}
Current evidence for dark matter in the universe does not exclude heavy composite nuclear-density objects
consisting of bound quarks or antiquarks over a significant range of masses. 
Here we analyze one such proposed scenario, which 
hypothesizes antiquark nuggets with a range of $B \sim 10^{24-30}$ with specific predictions for
spectral emissivity via interactions with normal matter. We find that, if these objects make
up the majority of the dark matter density in the solar neighborhood, their radiation
efficiency in solids is marginally constrained, due to limits from the total
geothermal energy budget of the Earth. At allowed radiation efficiencies, the
number density of such objects can be constrained to be well below dark matter densities 
by existing radio data over a mass range currently not restricted by other methods.
\end{abstract}
\maketitle

\section{Introduction}
Many different forms of astrophysical and cosmological evidence point to the existence of 
weakly interacting matter in an unknown form in the universe (see ~\cite{DM0} for a recent review), 
with a total mass of order five times the inventory of what is
observed as normal matter -- gas, stars, and dust -- giving a dark matter density
estimated to be in the range $0.4-1$~GeV~cm$^{-3}$ in the solar neighborhood~\cite{DM1,DM2,DM3}
In addition, the current best models for  the early universe give stringent constraints on the 
content of interacting baryons in this dark matter at the time of nucleosynthesis~\cite{DM0}. 

The rather compelling arguments for non-baryonic dark matter have led to a wide variety of efforts,
both theoretical and experimental, 
to either postulate or directly detect new particles, beyond the standard model, 
that would satisfy the dark matter characteristics. 
However, there remain several ``standard-model'' candidates for dark matter, which, if not currently
favored, have not yet been excluded over all the possible range of parameters. In particular,
very massive objects (at least compared to the $\lesssim$TeV scale of typical particle candidates for dark matter)
can still satisfy the astrophysical constraints on dark matter if their masses are sufficient 
that the flux in typical detectors is extremely low, but not so large that they are excluded due to
galaxy dynamics or gravitational lensing observations. This may be translated into a constraint
on their interaction cross sections per unit mass; 
current limits require $\sigma/M \lesssim 0.1$~cm$^2$~g$^{-1}$~\cite{CSS08,Peter12}, a constraint
that is in general easily satisfied by neutral objects with nuclear densities.

Such objects must not interact with normal matter via strong or electromagnetic channels 
at the time of nucleosynthesis. One candidate is the so-called quark nugget,
or strangelet, hypothesized originally by Witten~\cite{Witten84, Itoh70}, and developed in many variations
since then. Quark nuggets can be neutral and metastable at their formation during the quantum chromodynamics 
phase transition of early-universe
evolution, and thus do not undergo significant further interactions at nucleosynthesis, therefore evading
the constraints on baryonic content. 

Although still a matter
for debate, the possibility of quark nugget {\it color superconductivity}~\cite{Wil98},
in which quarks near the Fermi surface of the nugget form correlated Cooper pairs,
favors their possible stability~\cite{LH04}. One of the more attractive aspects of these objects as candidates for
dark matter is that the physics of their formation and interactions is in principle calculable according
to the standard model, although such calculations can in practice be prohibitively difficult.

\section{Antiquark nuggets}

A recent novel application of this model~\cite{FZ08,FLZ10,Lawson11, Lawson12}
postulates that both antiquark and quark nuggets are formed in the early universe with a ratio of 3:2,
and the current observed baryon matter-antimatter asymmetry arises only because the antibaryons are
hidden in the excess of antiquark nuggets (AQN), which, along with the quark nuggets form the
bulk of the dark matter. AQN have the same kinetic energy as normal quark nuggets, and the transfer of
this energy may be observed
in seismic~\cite{Seis1} or thermal events produced in the Earth's crust. However, in addition to kinetic energy
transfer,  AQN sweep up and annihilate with normal matter along their track, leading to potentially much more energetic
signatures and much higher rates of radiative energy deposition. 

Current constraints from seismic energy deposition
in the Earth and Moon indicate that quark nuggets can only satisfy dark matter density for baryon numbers
$B \lesssim 7 \times 10^{28}$~\cite{Seis3} (but see the possible detection of an event in ref.~\cite{Seis2}, and
the response in~\cite{Seis4}). Limits from non-detection of compatible events in the Lake Baikal
detector~\cite{Zhit03} require $B \gtrsim 1.2 \times 10^{20}$.

Taking an approximate geometric mean value of these constraints, a baryon number of $B \gtrsim 10^{24}$ ($\sim 1.6$ gm)
gives a flux of AQN at Earth, assuming they are virialized with Galactic velocities of order 200 km s$^{-1}$, 
of order several per km$^2$ per year if all AQNs were close to this mass; actual fluxes will depend of course on the 
assumed mass spectrum in the solar neighborhood. 

\subsection{AQN thermal emission}
\begin{figure}[htb!]
\includegraphics[width=3.4in]{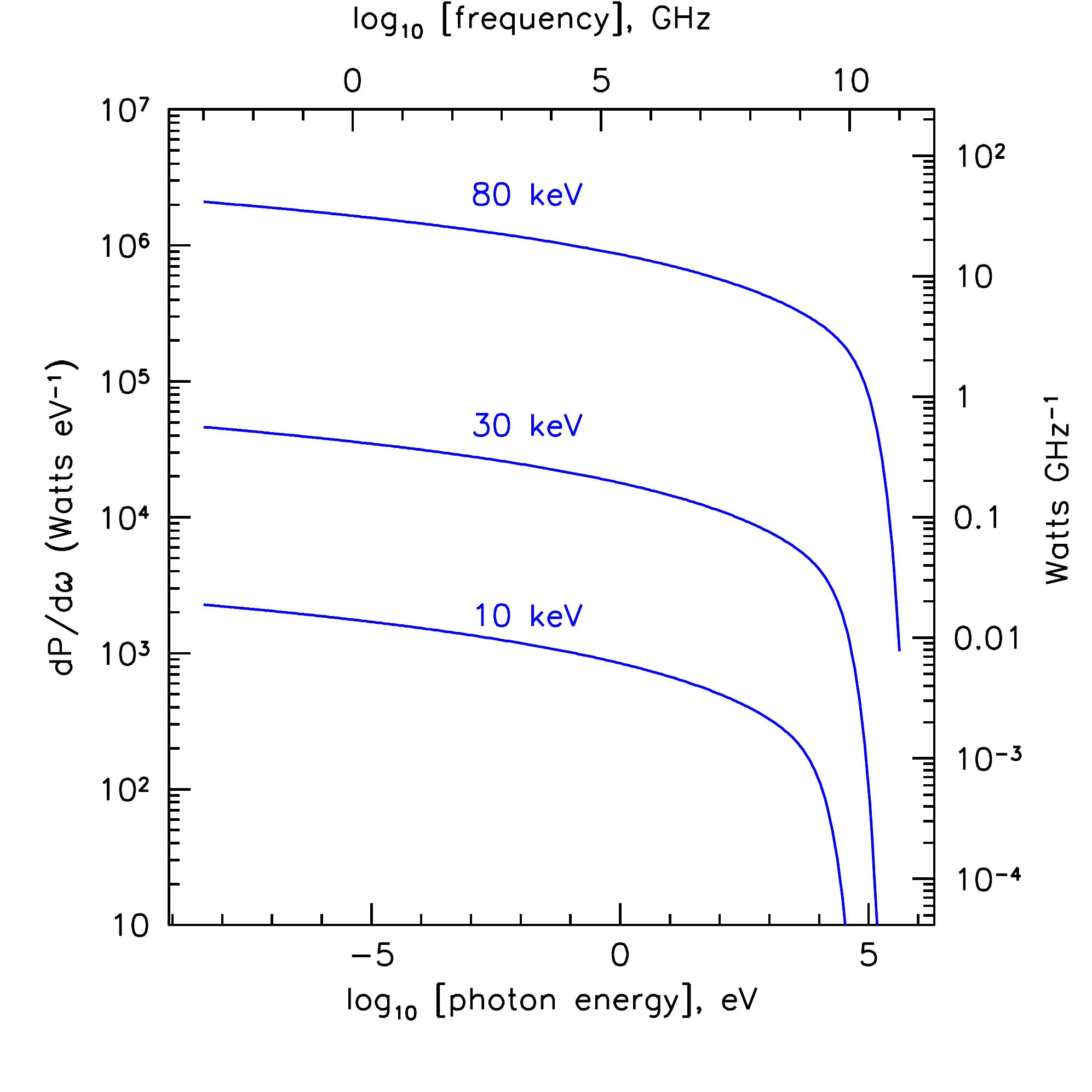}
\caption{\small \it  Spectral density of emission from a $B\sim 10^{24}$ antiquark nugget for three values of
the effective surface temperature.
\label{qnspectrum}}
\end{figure}

It is instructive to consider the flux of AQN at
this mass scale to determine the rate of energy deposited in the Earth. Recent detailed calculations
of the emissivity of AQN when accreting normal matter have been carried out using a Thomas-Fermi model~\cite{FZ08,FLZ10}.
In this case the spectral emissivity of a nugget at effective temperature $T$
for photon energies well below the electron mass $m_e$ is given by:
\beq
\label{emission}
\frac{dF}{d\omega} ~\sim ~ \frac{4}{45} \frac{\sigma_{SB} T^3 \alpha^{5/2}}{\pi} ~ 
\left ( \frac{T}{m_{e}} \right )^{\frac{1}{4}} ~ K\left ( \frac{\omega}{T} \right )
\eeq
where $\sigma_{SB}$ is the Stefan-Boltzmann constant, $\alpha$ is the fine-structure constant, and
$\omega$ is the angular frequency of the radiation, and
$$K\left ( \frac{\omega}{T} \right ) = \left ( 1 + \frac{\omega}{T} \right )  \left ( 17 - 12 \log\left [ \frac{\omega}{2T} \right ] \right ) ~e^{-\omega/T}~.$$
(Here ratios such as $\omega/T$ are given with
implicit values of Planck's and Boltzmann's constants $h,~k$ where necessary to rationalize the units.)
This result, converted to spectral power $dP/d\omega = 4\pi R_n^2 dF/d\omega$, is plotted  in Fig.~\ref{qnspectrum}
for the case of an AQN of $B = 10^{24}$, and  
a nugget radius estimated~\cite{FZ08,FLZ10} to be $R_n = 10^{-7}$~m,
for three different AQN surface temperatures.

The integral emissivity over all frequencies, determined from this spectral emissivity is
\beq
F_{tot} = \frac{d^2E}{dtdA} ~=~ \frac{16}{3}~\frac{\sigma_{SB} T^4 \alpha^{5/2}}{\pi} \left [ \frac{T}{m_e} \right ]^{\frac{1}{4}}
\label{ftot}
\eeq
and is typically a fraction of order $10^{-6}$ of blackbody emission~\cite{Lawson11,FZ08}. This power is however
distributed with a spectral emissivity very different from a blackbody, 
since it is nearly flat at low frequencies as equation~\ref{emission} and
Fig.~\ref{qnspectrum} show.

\subsection{Temperature upper bound}

The temperature of the AQN in equation~\ref{emission} is determined from the amount of 
matter accreted and annihilated along the nugget's path, by normalizing the total emission (the integral
of this equation over all $\omega$) to equal the fraction $1-g$ of annihilation energy that the
nugget thermalizes from its accretion~\cite{FZ08}, thus $F_{tot} ~=~ (1-g) F_{ann}$. For atmospheric
and surface densities encountered at Earth, the density-dependent effective temperature is expected to be~\cite{FZ08}
\beq
T_{\rho}=90~{\rm keV}\left [ \left ( \frac{1-g}{0.9} \right ) \left ( \frac{f}{0.067} \right ) \left ( \frac{u_n}{10^{-3}~c} \right ) 
\left ( \frac{\rho}{{\rm g~cm^{-3}}} \right ) \right ]^{\frac{4}{17}}
\label{TkeV}
\eeq
where $f$ is related to the accretion efficiency of the AQN, $u_n$ is the AQN velocity relative to Earth, and $\rho$ is the density
of the medium. The annihilation efficiency terms are somewhat uncertain; 
typical values given are $g\sim 0.1$ and $f \sim 0.067$~\cite{FZ08}.

A firm upper bound on the maximum AQN temperature $T_{max}$ comes from the requirement that 
the radiation pressure just outside the surface of the AQN should not exceed the level required to accelerate
the accreting matter away from the path of the nugget.
A previous qualitative analysis~\cite{Lawson12} of the effects of radiation pressure of incoming material in the Earth's atmosphere 
estimated that $T$ would saturate at densities $\rho \sim 10^{-3}$~g~cm$^{-3}$ in the lower atmosphere, giving $T_{max} \sim 10$~keV.
This estimate did not account for the likely photo-ionization of the region very near the nugget and thus
probably underestimated $T_{max}$~\cite{Lawson_pers}, since the radiation pressure primarily depends on the cross section
for photon momentum transfer with incoming nuclei. Another possible accretion-limiting effect was also discussed in reference~\cite{Lawson12},
that of momentum transfer to the incoming nuclei via scattering off positrons in the AQN electrosphere
surrounding the nugget. We first discuss the radiation pressure bound, the return later to the effects
of positron scattering.

To illustrate these effects, consider the case
of a nugget with $B = 10^{24}$, and $T \sim 10$~keV ($\sim 10^8$~K) accreting material at 
sea-level Earth-atmospheric density $\rho \sim 10^{-3}$g~cm$^{-3}$ at a velocity of 200~km s$^{-1}$. 
For this case, the AQN luminosity is $L_{tot} = 4\pi R_n^2 F_{tot} = 2 \times 10^6$~W. The intensity 
just outside the surface of the nugget is $I_{surf} = F_{tot} \sim 1.6 \times 10^{19}$~W~m$^{-2}$. 
Since a large fraction of this intensity is emitted as soft X-rays, for which the photo-ionization
cross section on air will far exceed the momentum-transfer cross section $\sigma_N$, the
incoming material will be fully ionized before it approaches the AQN surface. The residual momentum-transfer cross section on the stripped nuclei
is not well-documented, but it cannot exceed the incoherent atomic scattering cross section at keV photon
energies, thus we estimate $\sigma_{N} \lesssim 10^{-2}~{\rm barn}~ = 10^{-30}$~m$^2$. The required acceleration to displace
the nuclei from the oncoming nuggets path is of order $a_{min} = \Delta u/\Delta t \sim u_n^2/R_n$,
and thus the radiation pressure is determined from
\beq
\frac{I_{surf}}{c} ~\leq~ \frac{m_N u_n^2}{\sigma_{N} R_n} 
\eeq
where $m_N \sim 2.2 \times 10^{-26}$~kg is the typical atomic mass, for nitrogen in this case.
This yields, for the example above, $I_{surf} \lesssim 3 \times 10^{30}$~W~m$^{-2}$, many orders
of magnitude above the the AGN radiance in this case. Bounding the surface emissivity with this
then bounds the AQN temperature:
\beq
\frac{16}{3}~\frac{\sigma_{SB} T^4 \alpha^{5/2}}{\pi} \left [ \frac{T}{m_e} \right ]^{\frac{1}{4}}
\lesssim \frac{m_N u_n^2 c}{\sigma_{N} R_n} 
\eeq
and since the AQN radius depends on the number of baryons $B$ as $R_n \sim 10^{-7} (B/10^{24})^{1/3}~{\rm m^2}$~\cite{FLZ10}:

\begin{multline}
T_{max}~ \lesssim~  4.5~{\rm MeV}~ \left [ \left ( \frac{u_n}{\rm 200~km~s^{-1}} \right )^2 \left ( \frac{m_N}{\rm 14~amu} \right ) \right. \\
\times \left.  \left ( \frac{\sigma_N}{\rm 10^{-30}~m^2} \right )^{-1} \left ( \frac{B}{10^{24}} \right )^{-\frac{1}{3}} \right ]^{\frac{4}{17}}~. 
\label{radpress}
\end{multline}
Of course, at several MeV AQN surface temperatures, external nuclear interactions and associated energy release will
become important, and the luminosity of the nugget will be so high that other limiting effects are
likely to obtain well before these temperatures are reached. 

Another potential limiting effect is that the ram pressure of the fraction of stripped
nuclei that mechanically collide with positrons in the electrosphere of the nugget will form
a bow shock to the propagating nugget, and this will limit the ingress of other matter. This
constraint was discussed briefly by Lawson~\cite{Lawson12}, where it was argued that once the kinetic temperature
of the positrons approached that of the incoming nuclei, they would begin to scatter off the positrons and a
negative feedback condition would obtain. However, the momentum transfer of the electrosphere to incoming 
nuclei is given by the Rutherford scattering cross section for positrons on the stripped nuclei.
As the nugget temperature increases, the Rutherford scattering cross section decreases
quadratically with the effective temperature 
$\sigma_R(T) \propto T^{-2}$. In contrast, the AQN electrosphere temperature increases only slowly with 
annihilation rate $ T \propto F_{ann}^{4/17}$~\cite{FZ08}, 
causing this process to decrease in efficiency in deflecting nuclei as
the accretion process increases. Thus it appears that positron scattering cannot produce a negative
feedback accretion-limiting condition.

A more likely source of negative feedback may come from backscattering at the quark-matter interface
of the nugget; the accretion efficiency term $f$ used above arises from this process~\cite{FZ08},
but it is unclear what the temperature dependence of this effect might be.
An estimate of this process is beyond our scope, so in what follows we assume that
AQN surface temperatures approaching 100~keV (a few percent of the radiation pressure $T_{max}$ derived
here) in solid materials are not excluded yet by any accretion constraints.


\section{AQN interacting at Earth}

\subsection{Lithosphere interactions}

We first consider interactions of these AQN in 
Earth's lithosphere. The kinetic energy loss for either quark or anitquark nuggets is given by~\cite{Seis1} 
\beq
\label{kinetic}
\frac{dE}{dx} ~=~ -A_n \rho(x) u_n^2
\eeq
where $A_n$ is the cross sectional area of the nugget, and $\rho_(x)$ is the density along the track $x$.
Using the average density $\rho(x) = \langle \rho \rangle$, 
\beq
\label{uofx}
u_n(x) = u_n(0) e^{-A_n\langle \rho \rangle x/M_n}~.
\eeq
If an antiquark nugget stops in the Earth through loss of kinetic energy, it will then subsequently
completely annihilate, and the energy deposition in that case will equal its remaining mass energy.
For AQN to remain viable dark matter candidates, the total power contribution through this
process to the Earth's thermal energy budget must not exceed that of known sources. 
The current geothermal energy budget of the Earth is $P_{geo} \sim 44 \pm 3$~TW \cite{JLM07,DD10},
and of order half of this must be attributable to radionuclide decay; the remainder is still
a subject for debate, although a major fraction must be residual heat from the gravitational
collapse of formation. If we allow that of order $P_{geo}/4 \sim 11$~TW of the current geothermal energy budget
could be available to external heating from AQN annihilation, then the rate of 
captured AQN must satisfy $dm/dt \leq P_{geo}/(4c^2) = 0.12$~grams/sec. At the current firm lower bound for
AQN baryon number $B_{min} \sim 10^{20} = 0.16$~mg, a flux of AQN at this mass, equal to the dark matter density, 
is of order $10^6$~s$^{-1}$ over the whole Earth, thus the capture probability must not exceed 
about $10^{-3}$ per nugget. The required flux to match the DM falls as $B^{-1}$, and the mass-energy
rises as $B$, so this constraint is constant with AQN mass. So far we have ignored the energy deposited
during the transit of the nugget; we return to that below.

If we require that no more than 0.1\% of all AQN lose enough kinetic energy via equation~\ref{kinetic}
to be captured by Earth, this translates into a requirement that the velocity attenuation in equation~\ref{uofx}
above can only fall below the escape velocity in 0.1\% of all AQN tracks. Taking $u_n(0) = 200$~km~s$^{-1}$,
and the Earth escape velocity $u_{esc} \sim 11$~km~s$^{-1}$, we evolve an initial Maxwell-Boltzmann velocity
distribution by assuming equation~\ref{uofx} above, with a mean travel distance of $\bar{x} = 4R_E/3$,
the mean chord distance through the Earth for random tracks~\cite{Berengut72,KK03}. By then requiring that
the cumulative evolved Maxwellian have no more than 0.1\% of its final velocities below $u_{esc}$,
we find the following constraint if all of the dark matter consists of AQN of mass equal to or greater than $B_{min}$:
\beq
B_{min} ~=~ 2.6 \times 10^{24}
\eeq
where we have used a mean density of $\langle \rho \rangle=5500$~kg~m$^{-3}$ for the Earth, 
and a nugget area  $A_n = \pi R_n^2$,  with $R_n = 10^{-7} (B/10^{24})^{1/3}$~m. It thus appears that
geothermal considerations rule out AQN of masses less than about 4 grams, well above prior constraints.
This limit can only be evaded if the velocity distributions of the AQN are decidedly non-virial,
but a similar constraint will obtain on whatever velocity distribution is present.

Now consider the emissivity in lithosphere transit of a flux of AQN well above $B_{min}$ given here, with a nearly
mono-mass spectrum with $B\sim 10^{25}$. As the nugget enters the solid crust at $200~$km~s$^{-1}$, the temperature
rises to around 120 keV (using expected values for $f,~g$ above, and a mean density of 5.5 gm~cm$^{-3}$, 
giving an initial luminosity $P_i = 4\pi~ R_n^2 ~F_{tot} \sim 3.7 \times 10^{11}$~W. At this mass, the velocity
attenuation length is $M/(A_n \langle \rho \rangle) \sim \pi R_E$, and
using the mean chord of $4R_E/3$, the velocity at exit is $u_{f} \sim (2/3)u_n(0)$, so the mean velocity is of
order $0.8 u_n(0)$, and thus the average power is $0.8 P_i \sim 3 \times  10^{11}$~W.
The integrated rate of energy being continuously deposited, if the dark matter consists entirely
of nuggets of this mass, is $P_{tot} \sim 48$~TW. This level of thermal energy also exceeds
the current $\sim 44$~TW geothermal energy budget of the Earth\cite{JLM07,DD10}.
Using again the requirement that AQN contribute no more than 1/4 of the current geothermal energy,
we may place a constraint on the maximum temperature at $B \sim 10^{25}$: 
in this case $T_{max} \leq 120~{\rm keV} (11/48)^{4/17} = 85$~keV. 

It might appear that the $T^4$ dependence of the AQN luminosity does not leave much headroom for larger 
nugget masses, since the accretion rate grows with the cross section of the nugget. Clearly,
the radiation pressure constraint used in deriving equation~\ref{radpress} above is far less restrictive than
the constraints from geothermal power. However, 
the temperature is to first order only dependent on the density of the medium, rather than nugget mass. 
Also, since the accretion cross section grows as $B^{2/3}$, 
and the flux required to match the dark matter density decreases as $B^{-1}$, higher masses are possible,
but again will be marginally close to violation of the geothermal constraints, unless the maximum
temperature or accretion efficiency is lower than initial estimates.

\subsection{AQN meteors?}

Propagating in the Earth's lower atmosphere at a temperature of 10~keV, 
an AQN with $B \sim 10^{24}$ produces megawatt total bolometric luminosity. However, its apparent visual signature in the optical 
band would not necessarily be dramatically different
than a very fast meteor. At $u_n \sim 200$~km~s$^{-1}$ it is only a factor of 3 faster than the fastest
meteors, although its trail could be much longer. At this temperature, the AQN power in the optical band is
of order 1 kW, equivalent to relatively bright meteor with visual magnitude of $m_{V} \sim -1$~\cite{Opik55}.
However, the nugget would not achieve this temperature until near ground level; at typical
$\sim 80$~km meteor altitudes, the AQN temperature would be a factor of 10 lower, and the luminosity
more like 100~mW, thus making it invisible to the naked-eye at higher altitudes, where it would be
much more widely visible. It is not hard to see why
such events could evade casual observation. 

\subsection{Thermal radio signatures}

While the visual emissivity of an AQN with $B \sim 10^{24}$ passing through the terrestrial atmosphere may be easily
missed, the flat spectrum displayed in Fig.~\ref{qnspectrum} produces a surprisingly strong broadband radio
signature, of order 10~mW GHz$^{-1}$ in the VHF to microwave band even at the lowest AQN temperature considered,
and far more at the higher ones. The expected radio flux density in this spectral region 
for an AQN transiting the atmosphere is thus of order~\cite{Lawson12}
\begin{multline}
\label{satm}
S_{AQN,atm} \sim 8 \times 10^{-23}~{\rm W~m^{-2}~Hz^{-1}}~\left ( \frac{T}{\rm 10~keV} \right )^{13/4} \\
\left ( \frac{\langle D\rangle}{\rm 100~km} \right )^{-2} \left ( \frac{B}{10^{24}} \right )^{2/3}
\end{multline}
where we have ignored the weak logarithmic dependence. For an AQN transiting a solid material,
viewed from outside the material, there are additional terms due to the attenuation of radio signals in the solid material,
and the Fresnel transmission coefficient $\mathcal{F}_T$ for the emission as it passes through the interface. Thus
\beq 
\label{ssolid}
S_{AQN,solid} ~=~ S_{AQN,atm}~\mathcal{F}_T~e^{-2 d / L_{atten}}
\eeq
where $d$ is the pathlength of the radio emission in the solid, and $L_{atten}$ is the field attenuation length in the medium.

Receiver thermal noise levels at a typical receiver system temperature of $T_{sys} \lesssim 300$K, by comparison, are typically 
$P_n = k T_{sys} \Delta f \lesssim 5$~pW in a GHz of bandwidth. The broadband AQN radio flux density is thus likely to be
well above thermal noise for large distances from the track. This will of course depend on the
mean distance $\langle D\rangle$ from the track, as well as the time $\tau$ over which the AQN track remains in the primary
field-of-view, or half-power beamwidth $H$, of a given antenna. This in turn depends on the antenna gain (or directivity) $G$.
For moderately directive antennas of a few dBi of gain or more, $G \sim  27000 (H^{\circ})^{-2}$ where the beamwidth $H$ is
given in degrees here, and thus for $u_n \sim 200$~km~s$^{-1}$, $G \sim 10$, and ${\langle D\rangle} \sim 100$~km,
the in-beam residence time is 0.3-0.5 seconds; however, receiver gain instabilities make it practical to limit 
the integration time $\tau \lesssim 0.1$ seconds, giving several samples per transit per beam. 
An antenna of constant gain $G$ over a passband from $f_1$ to $f_2$ 
has an average effective area (for a flat spectrum source) of $A_{eff} = 2 G c^2 /(4 \pi f_1 f_2)$
and the minimum detectable signal power of this antenna with 
receiver bandwidth $\Delta f = f_2-f_1$ and integration time $\tau$ is
\beq 
\sigma_{s} ~=~ \frac{kT_{sys}}{A_{eff} \sqrt{\Delta f \tau}}~.
\eeq
Assuming the integration time $\tau$ is matched to the expected beam crossing time 
for $u_n \sim 200$~km~s$^{-1}$,  the limiting sensitivity is

\begin{multline}
\label{sigmas}
\sigma_{s} \simeq 10^{-24}~{\rm W~m^{-2}~Hz^{-1}}~~\left ( \frac{T_{sys}}{\rm 300~K} \right ) \left ( \frac{\sqrt{f_1 f_2}}{\rm 490~MHz} \right )^{2}\\
\times \left ( \frac{\Delta f}{\rm 1~GHz} \right )^{-1/2}
\left ( \frac{G}{\rm 10} \right )^{-3/4} \left ( \frac{\langle D\rangle}{\rm 100~km} \right )^{-1/2}~.
\end{multline}
Comparing this to equations~\ref{satm} and \ref{ssolid} 
indicates that such events are detectable with a modest antenna collecting area and receiver
out to distances of several hundred km, even at the lowest AQN temperature considered here. 
The advantage in detection of thermal radio emission as compared to other possible forms of beamed emission,
such as geosynchrotron emission considered in reference~\cite{Lawson12}, is that the acceptance solid
angle does not require observation very close to the axis of the AQN velocity. Thus for an isotropic
flux of AQN, thermal radio detection will have a far greater probability.

\begin{figure}[htb!]
\includegraphics[width=3.4in]{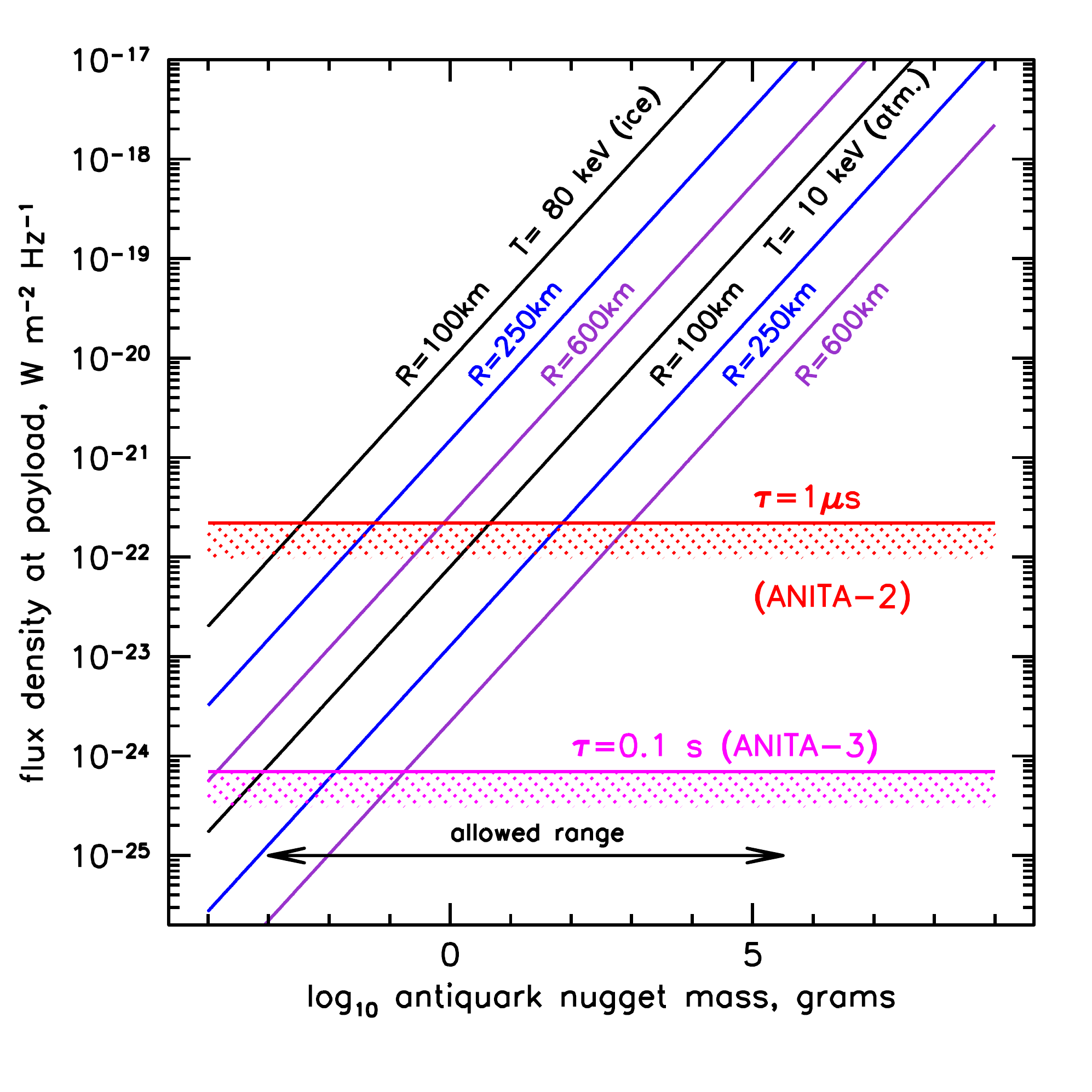}
\caption{\small \it Estimated flux density and sensitivity for ANITA for two different AQN temperatures
and three different distances, for AQN transiting in atmosphere ($T=10$~keV), or 
Antarctic ice ($T= 80$~keV), with typical estimated
attenuation losses and Fresnel coefficient. The thermal noise levels for two different integration
times are shown.
\label{QN1}}
\end{figure}

\section{Thermal radio detection prospects}

Given that the distance scale for detection is plausibly hundreds of kilometers even at the lowest AQN
temperatures, and perhaps much more at higher temperatures, it is evident that a ground-based detector
is at a disadvantage compared to the synoptic field-of-regard of suborbital or orbital platforms. 
Since the AQN luminosity appears to rise by orders of magnitude once it enters solid material, 
and since terrestrial ice in the Earth's cryosphere is highly transparent in the VHF and UHF radio range~\cite{icepaper,Ice09},
suborbital/orbital observations of Antarctic or Arctic ice sheets afford perhaps the most sensitive possible 
channel for AQN detection. We thus conclude by estimating to first-order the sensitivity of this approach,
using the parameters of the Antarctic Impulsive Transient Antenna (ANITA) suborbital payload~\cite{ANITA-inst}, 
which has completed two flights (ANITA-1: 2006-2007; ANITA-2: 2009-2010) and is scheduled to complete a third next year.
ANITA has enough directional capability to establish the velocity of an AQN candidate, a crucial discriminator against
other possible background events such as meteors.

For a radio detector viewing Antarctica synoptically from stratospheric altitudes, as in the case of ANITA,
the horizon is at a distance of $600-700$~km, and the area viewed is over 1M km$^2$ out to the horizon. To
illustrate the range of sensitivity, Fig.~\ref{QN1} plots the AQN signal and thermal noise levels
based on equations~\ref{satm},~\ref{ssolid}, \&~\ref{sigmas} above, using instrument parameters for ANITA-2 and ANITA-3~\cite{ANITAparam},
for a range of AQN masses that are currently unconstrained, for several different distances. For ANITA-2,
the $\tau=1~\mu$s integration time arose in the ambient radio-frequency (RF) power monitoring system, an auxiliary
detector to the primary 2.6 Gsample/sec waveform recorder which captures only a 100~ns time window. The
ambient RF power monitor samples each antenna signal at about 8~Hz, but with a much shorter integration
time due to analog-to-digital conversion related constraints. However, these samples occur for both polarizations,
and there are 2-3 antennas sampling each azimuthal direction at $22.5^{\circ}$ azimuth intervals.
Since the thermal AQN signals are unpolarized, the two ANITA polarization samples are independent, as are
the 2-3 different antennas per azimuthal sector, and thus an AQN signature can be detected with high-confidence
by requiring the signals to appear in a majority coincidence of all of these independent channels. For
ANITA-3 the design will allow for much longer integration times per RF power monitoring signal, and thus
the sensitivity will be substantially improved.
    
It is evident from Fig.~\ref{QN1} that AQNs transiting either the atmosphere or 
ice sheets can be detected, but to ensure that these are distinguished
from the many forms of anthropogenic and other radio-frequency interference, the AQN track needs to be detected over
an azimuth range such that a velocity can be unambiguously determined. This requires a projected azimuthal span for the track of order
$\Delta \phi \gtrsim 20^{\circ}$ for ANITA~\cite{ANITA-inst}, so that at least two adjacent azimuthal sectors show a signal;
this is possible over this range of azimuth since the ANITA antenna beams overlap each other in adjacent azimuthal sectors.

\begin{figure}[htb!]
\includegraphics[width=3.4in]{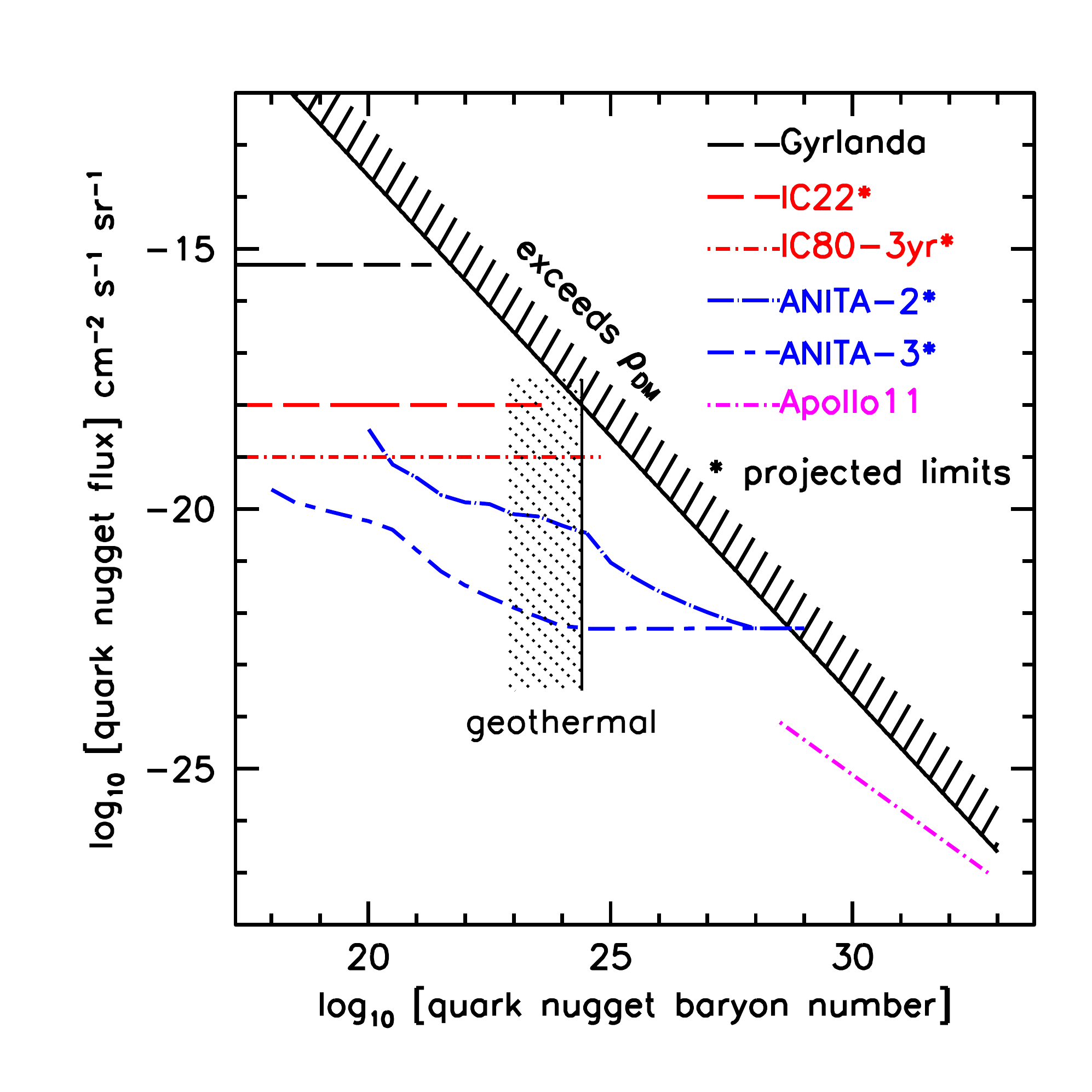}
\caption{\small \it Existing and projected limits on quark nugget fluxes with respect to dark matter density 
(diagonal hatched curve). The ANITA curves are estimates from this work.
\label{QNlimits}}
\end{figure}

To estimate the limiting sensitivity for ANITA, we must integrate over the acceptance area, solid angle, and
detection efficiency. The number of detected events $N$ as a function of
the baryon number $B$ of the AQN can be written
\beq
\frac{dN}{dB} ~=~ T_{obs} \int_0^{4\pi} d\Omega \int_0^{R_{H}} 2 \pi r dr \mathcal{F}(B) \mathcal{E}(B,r,\theta,\phi,\rho) 
\eeq
where $T_{obs}$ is the integrated observation time, 
$R_H$ is the distance to the horizon, $\mathcal{F}(B)$ is the flux of AQN, and $\mathcal{E}(B,r,\theta,\phi,\rho)$
is the instrumental detection efficiency (bounded between [0,1]) as a function of baryon number, 
distance $r$, angular track directions $\theta,\phi$, and
medium density $\rho$.

The requirement on azimuth span above constrains the track to be relatively horizontal,
such that at distance $D$ from the payload, the zenith angle range is limited by $\Delta \theta \sim (10~{\rm km})/(D sin \Delta \phi)$ for
events in the atmosphere, $\Delta \theta \sim (2.0~{\rm km})/(D sin \Delta \phi ) $ for in-ice tracks.  
Using these constraints, and requiring that the signal-to-ratios 
$S_{AQN,atmos}/\sigma_s ~\geq ~3$ or $S_{AQN,ice}/\sigma_s ~\geq~3$ for a trigger, we have simulated
the detection of AQNs with ANITA to determine the limiting sensitivity using Monte Carlo methods.
The results are shown in Fig.~\ref{QNlimits} where we have shown the curves based on the flown ANITA-2, and planned ANITA-3 instruments.
For ANITA-2, this applies to data that has been already acquired, but has not yet been analyzed for this type of signal; 
here we assume that no background events are found.
For ANITA-3 we use projected performance estimates provided by the ANITA collaboration, and a 30 day livetime is assumed,
similar to what was achieved for ANITA-2. The inflection in the sensitivity curve in each case is due to the 
crossover of the effects of ice detection, which is more efficient at lower AQN masses, and atmospheric detection,
which has a larger available solid angle and is more effective at higher AQN masses. For both ANITA-2 and ANITA-3,
the sensitivity eventually saturates the available area for very high AQN masses, and the curves flatten out.

We have included in Fig.~\ref{QNlimits} the constraint from geothermal power derived above, 
and we note the other two actual (rather than projected) limits 
are from the Lake Baikal detector~\cite{Zhit03}, and from analysis of the Lunar seismic noise detected in the Apollo 11
mission~\cite{Seis3}. For the IceCube curves, both are still projections, although the data for IceCube 22 already exists. For IceCube 80,
the projection is for three years of data, which will be achieved in early 2014~\cite{IceCubeExotics}. 
Analysis of existing ANITA-2 data for these event signatures has begun, and results may be expected within the next year.
Thus it appears that this very interesting region of parameter space for quark nugget dark matter is within reach of
several current experiments, and we can hope for either detections or compelling limits in the near future.

We thank both NASA and the US Department of Energy, High Energy Physics Division for their generous support of this work, and
Kyle Lawson for very useful input. We also thank the ANITA collaboration for providing information regarding
ANITA-3 performance, especially Gary Varner and Patrick Allison for their help in understanding the existing
instrument performance.


\begin{thebibliography}{99}
\bibitem{DM3} The actual value of the local dark matter density remains a topic of 
lively debate, with some estimates as low as zero, to values about two to three times that
quoted here.
\end{thebibliography}
\end{document}